\documentstyle[tighten,preprint,eqsecnum,aps,prd]{revtex}
\begin{document}
\draft
\setlength{\baselineskip}{16pt}

\def\S{Schwarzschild\enspace}
\def\l{\Lambda}
\def\P{{\tilde P}_R}
\preprint{\vbox{\baselineskip=35pt
\rightline{JYFL Preprint 18/1998}
\rightline{\ \ }
\rightline{\ \ }
\rightline{\ \ }
}}

\title{How to interpret black hole entropy?}

\author{Jarmo M\"akel\"a\footnote{Electronic address: Jarmo.Makela@phys.jyu.fi} and
Pasi Repo\footnote{Electronic address: Pasi.Repo@phys.jyu.fi}}
\address{Department of Physics, University of Jyv\"askyl\"a, P.O. Box 35, SF-40351
Jyv\"askyl\"a, Finland}

\maketitle

\begin{abstract}
We consider a possibility that the entropy of a \S black hole has two 
different interpretations: The black hole entropy can be understood either as an outcome
of a huge degeneracy in the mass eigenstates of the hole, or as a consequence of the fact 
that the interior region of black hole spacetime is separated from the exterior region by a
horizon. 
In the latter case, no degeneracy in the mass eigenstates needs to
be assumed. Our investigation is based on calculations performed with Lorentzian 
partition functions obtained for a whole maximally extended \S spacetime, and for its
right-hand-side exterior region. To check the correctness of our analysis we  
reproduce, in the leading order approximation, the Bekenstein--Hawking entropy of the
\S black hole.
\end{abstract}


\narrowtext

%\twocolumn%

\section{Introduction}
\label{sec:intro}

One of the most interesting branches of modern theoretical physics is black hole thermodynamics.
The origin of this fascinating area of research can be traced back to the early 70's when it
was observed that there are certain striking similarities between the laws of black hole mechanics
and the laws of thermodynamics\cite{1,2}. The similarities 
were mostly considered artificial until Hawking convincingly found -- by taking into account 
quantum mechanical effects -- that the exterior region of a black hole produces thermal 
radiation\cite{3}. Ever since, the thermodynamical properties, like entropy, of black holes 
have been studied seriously, but many unsolved problems are still waiting for solutions. 
One of the most interesting issues is the question of the underlying microstates of the hole itself\cite{4}. The 
unknown microstates determine the average values of the thermodynamical quantities of the 
hole, and it is very likely that the solution of the problem of the underlying 
microstates of the hole will give us valuable clues to the self-consistent quantum theory of 
gravity.

We all have been convinced by now by the fact that black holes bear entropy $S=\frac{1}{4}A$\cite{2}.
This result originates from Bekenstein's and Hawking's work\cite{1,3}, but it has been reproduced
by many authors since then\cite{5,CentSur,6,7,8,9,10,11}. The original calculation yielding the entropy was based 
on the semiclassical gravity, where spacetime was considered as a classical object, whereas
matter fields were quantized in this classical but curved background spacetime. Some years 
later, Hawking was able to recover the same result by means of a Euclidean path-integral approach to 
quantum gravity\cite{CentSur}. Those approaches, 
however, failed to give an explanation to the black hole entropy at the fundamental level. More
precisely, they did not provide a solution to the problem of underlying microstates of the 
hole: Since the black hole entropy is $\frac{1}{4}A$, one might expect that there are
$\exp(\frac{1}{4}A)$ microstates corresponding to the same macrostate of the hole, and the problem 
is to identify these microstates. Search for the microstates has been going on for almost 
thirty years, and only recently the string-theoretical work of Strominger and Vafa has been 
able to give explicitly the number of the microstates\cite{12}. In this paper, however, we shall 
investigate the black hole entropy by means of canonical methods.

The classical no-hair theorem states that after the collapse, when a black hole has settled
down to a stationary state, its properties are determined by very few parameters 
observed far from the hole: These parameters are the mass $M$, the charge $Q$ and the angular momentum $\vec J$ of the
hole\cite{13}. Thus, from the classical point of view, black holes have only three degrees of freedom. 
What has happened to the enormous amount of degrees of freedom of the collapsing matter? The no-hair theorem prompts one
to believe that these degrees of freedom, and the information contained in them, is lost in the collapse, and
that the entropy of a black hole may be understood as a 
measure of information loss during the gravitational collapse, because between the entropy and 
the information there is a well-known relationship given by Brilliouin\cite{14}: the decrease in
information increases the entropy. This viewpoint is purely quantum-mechanical. According to 
quantum mechanics all the information from the collapsing star is not able to reach to an 
observer exterior to the newly formed event horizon.
In other words, all the microstates of the collapsing star cannot be measured by the 
external observer. This results to an increasing entropy $S$. 

The question now arises: After the collapse of
matter, are the degrees of freedom contained in the matter fields somehow encoded into the quantum
states of the black hole spacetime itself, or have they vanished altogether, leaving no trace whatsoever? Of 
course, it is natural to claim that they are encoded into the quantum states of spacetime itself such that
there is a vast $\exp(\frac{1}{4}A)$-fold degeneracy in the quantum states 
of the hole. This leads us to a conclusion that the total number of unknown 
quantum states of the black hole must be enormous, too. Thus, from a quantum-mechanical 
point of view, the number of the physical degrees of freedom of the hole is not limited to just few 
parameters. The contradiction between quantum and classical black holes is obvious: The number of
physical degrees of freedom of the classical hole is three, whereas the number of physical degrees of
freedom of the quantum black hole is enormous. The problem with this contradiction is that it is not quite clear how,
starting from general relativity, quantization itself might bring along a huge number of additional degrees of freedom.

The purpose of this paper is to investigate a possibility that the entropy of a black hole is
reproducable from the point of view of an external observer even if the observer takes 
into account the classical degrees of freedom only, and quantizes all the classically 
observed quantities, like mass, charge and angular momentum without assuming any degeneracy in the
eigenstates of these quantities. In other words, we shall consider a possibility that the other but the classical
degrees of freedom 
associated with the collapsing matter fields have vanished altogether. This point of view might provide a
solution to the apparent contradiction related to the number of degrees of freedom of quantum and classical
black holes. The
key point in this paper is that we investigate the statistical mechanics of the {\it exterior} region of the
black hole spacetime. This kind of a choice may be considered justified on grounds of the fact that the
interior region of the black hole is separated from the exterior region by a horizon. Hence, an external
observer cannot make any observations on the interior region, and one is justified to take a point of view that, 
for such an observer, physics of a black hole is physics of its exterior region. 
For the sake of convenience, we shall consider static and vacuum black holes only, but an analogous 
treatment could be performed for static electrovacuum black holes as well. The uniqueness 
theorem for nonrotating and vacuum black holes states that the \S metric, with the mass parameter 
$M$, represents the only static and asymptotically flat black hole solution\cite{13}.
We shall see that the Bekenstein--Hawking entropy of a black hole is reproducable from the statistical 
mechanics of the exterior region of \S black hole spacetime,
even if we assume that there is no degeneracy in the mass eigenstates of the hole. We shall also see that
the Bekenstein--Hawking entropy can be obtained for the whole spacetime as well, but in that case we
must assume, {\it a priori}, an $\exp ( \frac{1}{4}A )$-fold degeneracy in the mass eigenstates.

The analysis performed in this paper is based on the so called 
Hamiltonian thermodynamics of black holes. This branch of physics is an outgrowth
of the analysis on the Hamiltonian dynamics of the \S spacetimes performed by Kucha\v r\cite{15},
and was initiated, among others, by Louko, Whiting, and Winters-Hilt\cite{16,17}. We want to 
emphasize that the whole analysis in this paper is performed in Lorentzian  
spacetime without euclideanizing neither the Hamiltonian nor the action. The reason for
performing the analysis in Lorentzian spacetime is that the interior of the \S black hole 
is included in the analysis, too. In contrast, when one performs the euclideanization of the 
\S spacetime action, the black hole interior is reduced to one point and thus it is 
somewhat questionable to talk about the quantum states of the hole. In our investigations
the interior, as well as the exterior region of the hole plays an essential role. 

This paper is organized in the following way: In Sec. \ref{sec:Hamtheory} we describe very briefly
the Hamiltonian formulation of \S spacetimes and represent the Hamiltonian produced by
Louko and Whiting\cite{16} for the exterior region of the \S black hole. In Sec.
\ref{sec:Hamthermo}
we write two Lorentzian partition functions for the \S black hole. The first of these
partition functions describes the whole Kruskal spacetime, 
and the second the exterior region of the hole from the point of view of an
observer at rest relative to the hole at the 
right-handed asymptotic infinity. These two partition functions appear to give identical partition functions 
for the radiaton emitted by the hole, if we use Bekenstein's proposal for a 
discrete area spectrum, and assume, in addition, that all the energy and the entropy of the hole 
is exactly converted into the energy and the entropy of the radiation. 

The point we try to 
emphasize is that in order to obtain the partition function describing the whole spacetime,
the observer must accept an $\exp ( \frac{1}{4}A )$-fold degeneracy in the energy eigenstates of the hole, whereas 
no degeneracy needs to be assumed when one writes the partition function describing the
exterior region of the hole. This will be the main result of this paper, and it has an interesting 
consequence: If one takes a view that, for an external observer, only the physical properties of the exterior 
region of the hole are relevant, then it is not necessary to consider the possible internal degrees of 
freedom of the hole itself, but it is sufficient to take into account only the classical physical degree of 
freedom of the \S black hole, namely the mass $M$, to obtain the Bekenstein-Hawking entropy. This result
is in harmony with the no-hair theorem and with the semiclassical results. Unless 
otherwise stated, we shall use natural units where $c=G=\hbar =k_B=1$.
 
\section{Hamiltonian Theory}
\label{sec:Hamtheory}

In this section we shall give a brief introduction to the classical Hamiltonian theory
of spacetimes containing a \S black hole. We have not aimed at a presentation that
would give a technically detailed review on the subject; for more information, the authors
recommend the reader to consult the papers written by Kucha\v r\cite{15}, and by Louko and
Whiting\cite{16}. The classical Hamiltonian theory presented in this section is based on 
those papers.

The first successful Hamiltonian formulation of general relativity was the
so called ADM-formalism, which was discovered by Arnowitt, Deser and Misner\cite{18}. The basic
idea of the ADM formalism is to foliate the spacetime manifold into the spacelike 
hypersurfaces where the time $t=constant$ and to use the components of the induced three-metric tensor
$q_{ab}$ as the coordinates of the configuration space. It is clear that the formalism
depends heavily on the foliability of the spacetime manifold.

The ADM formalism of general relativity has four constraints per 
spacetime point, namely the Hamiltonian constraint and three diffeomorphism constraints.
The three diffeomorphism constraints imply an invariance of general relativity 
under spacelike diffeomorphisms, and the remaining Hamiltonian constraint implies 
an invariance in time reparametrizations. In addition to these four constraints, the 
formalism has, of course, the Hamiltonian equations of motions. These equations plus 
the constraints of the Hamiltonian theory are equivalent to Einstein's field equations
of general relativity.

When quantizing gravity canonically, we have to choose between two different 
possibilities: we either solve the constraints at the classical 
level, identify the physical degrees of freedom of the system and quantize the theory 
in the physical phase space, or we solve the quantum counterparts of the classical 
constraints. The former quantization method is known as the reduced phase space 
quantization, whereas the latter is known as the Dirac quantization\cite{19}. In this paper we 
shall use the results based on the reduced phase space formalism. The quantization 
of the physical degrees of freedom of the system will not be performed explicitly. 
Quantum theories of the \S black hole in the reduced phase space formalism have been 
constructed, among others, by Kucha\v r\cite{15} and by Louko and M\"akel\"a\cite{20}.

The classical constraints for spherically symmetric, asymptotically flat vacuum 
spacetimes have been solved, among others, by Kucha\v r\cite{15}, and by Thiemann and Kastrup\cite{21}. 
The only spherically 
symmetric, asymptotically flat vacuum solution to Einstein's field equations is the \S
solution. When the spacelike hypersurfaces, where $t=constant$, were chosen to go from 
the left to the right asymptotic infinities in the Kruskal diagram, crossing both the 
horizons, and the constraints were solved, Kucha\v r found that only two canonical 
degrees of freedom are left. If these two degrees of freedom are chosen to be the \S 
mass $m$, and its conjugate momentum $p_m$, the classical action of the system is
\begin{equation}
S_{\rm K} = \int dt \left[ p_m {\dot m} - m \left( N_++N_- \right) \right] \ \ ,
\label{Kuchact}
\end{equation}
where $N_+$ and $N_-$, respectively, are the lapse functions at the right and at the left
asymptotic infinities in the Kruskal diagram. The classical Hamiltonian of the whole 
maximally extended \S black hole spacetime found by Kucha\v r can therefore be written 
in terms of the two physical phase space coordinates $m$ and $p_m$ as:
\begin{equation}
H_{\rm whole} = m \left( N_++N_- \right) \ \ .
\label{Hwhole}
\end{equation}
  
The classical Hamiltonian theory of the right-hand-side exterior region of the \S black 
hole was investigated by Louko and Whiting\cite{16}. It follows from the analysis performed by those 
authors that, in the reduced phase space formalism, the classical  Hamiltonian describing
such a region of black hole spacetime can be written in terms of the \S mass $m$ and its
conjugate momentum $p_m$ as: 
\begin{equation}
H_{\rm ext} = mN_+-\frac{1}{2}R_h^2N_0 \ \ ,
\label{Hext}
\end{equation}
where $R_h=2m$ is the \S radius, $N_0$ is a function of the global time $t$ at the bifurcation
two-sphere such that
\begin{equation}
\Theta := \int_{t_1}^{t_2} dt N_0 (t)
\label{boost}
\end{equation}
is the boost parameter elapsed at the bifurcation two-sphere during the time interval
$[t_1,t_2]$, and, as before, $N_+$ is the lapse function at the right-hand-side asymptotic
infinity. We shall now give a brief review on the analysis performed by Louko and Whiting to 
produce the Hamiltonian (\ref{Hext}). 

Louko and Whiting considered a spacetime foliation where the spacelike hypersurfaces 
begin from the bifurcation two-sphere, and end at a right-hand-side timelike 
three-surface, i.e. at a "box wall" in the Kruskal diagram. With this choice, the spatial slices
are entirely contained within the right-hand-side exterior region at the Kruskal spacetime.
One of the main observations was that such foliations bring along an additional boundary term
into the classical action. Hence, the Louko-Whiting boundary action $S_{\partial \Sigma}$ consists of terms resulting
from the initial and the final spatial surfaces, that is, from the bifurcation two-sphere
and from the "box wall". After solving the classical constraints, Louko and Whiting found that
when the physical degrees of freedom are identified, the true Hamiltonian action is
\begin{equation}
S_{\rm LW} = \int dt \left( p_m{\dot m} - h(t) \right) \ \ ,
\label{LWact}
\end{equation}
where $h(t)$ is the reduced Hamiltonian such that, when the radius of the initial 
boundary two-sphere does not change in time $t$, the Hamiltonian $h(t)$
is defined as
\begin{equation}
h(t) := \left( 1-\sqrt{1-\frac{2m}{R}} \right) R \sqrt{-g_{tt}} - 2N_0(t) m^2 \ \ ,
\label{h}
\end{equation}
where $R$ is the time independent value of the radial coordinate of general spherically symmetric, asymptotically
flat vacuum spacetime at the final timelike boundary i.e. at the "box wall", and $g_{tt}$
is the $tt-$component of
the metric tensor expressed as a function of the canonical variables after performing 
a canonical transformation, and of Lagrange's multipliers. Details can be seen in Ref.\cite{16}.
It is easy to see that if one transfers the "box wall" to the asymptotic infinity by taking 
the limit $R\rightarrow\infty$, the Hamiltonian $h(t)$ of Eq. (\ref{h}) reduces to the
Hamiltonian $H_{\rm ext}$ of Eq. (\ref{Hext}). 
 
\section{Hamiltonian Thermodynamics}
\label{sec:Hamthermo}

If $\hat H$ is the Lorentzian Hamiltonian operator of a system, the partition 
function of the system is
\begin{equation}
Z=Tr \exp (-\beta \hat H) \ , 
\label{defpart}
\end{equation}
where $\beta = (k_B T)^{-1}$, $k_B$ is Boltzmann's constant and $T$ is the temperature of 
the system in a thermal equilibrium. The partition function (\ref{defpart}) corresponds to 
the canonical ensemble and describes the thermodynamics of the system in a thermal equilibrium.
Black holes can be considered as thermodynamical objects in a heat bath of temperature $T$\cite{2,5,22,23}.  Therefore,
if the system under consideration is the whole maximally extended \S spacetime, its Lorentzian Hamiltonian 
operator $\hat H$ would yield, via Eq. (\ref{defpart}), a non-Euclideanized thermodynamical description of the 
whole black hole spacetime, and if the system under consideration is the exterior region of 
the \S black hole only, the Lorentzian $\hat H$ would yield a non-Euclideanized partition function corresponding 
to the thermodynamical properties of the exterior region of the black hole spacetime. In practice, 
when one calculates the partition function (\ref{defpart}) one needs to know, or assume, the density 
of the energy states of the system. We shall come to this crucial point later on this section.

We first obtain the partition function corresponding to the whole maximally 
extended \S spacetime. Classically, $H_{\rm whole}$ may be understood as the total energy of
the whole spacetime. To choose a specific observer, who measures the energy of the gravitational field, 
we fix the values of the lapse functions at asymptotic infinities. From the point of view of an 
observer at the right-hand-side infinity at rest with respect to the hole, we can set $N_- = 0$ and 
$N_+ = 1$.  In other words, we have chosen the time coordinate at the right infinity to be the 
proper time of our observer and we have "frozen" the time evolution at the left infinity. The physical
justification for such a choice is that our observer can make observations at just one asymptotic
infinity. On the other hand, 
one may view the \S mass $m$ as the total energy of the \S spacetime, measured by the distant 
observer. Hence, we may write $H_{\rm whole} = m$.

To obtain the partition function for the Kruskal spacetime, we have to replace the operator 
$\hat H$ in Eq. (\ref{defpart}) by an operator counterpart $\hat H_{\rm whole}$ of the Hamiltonian
$H_{\rm whole}$. Hence, we get:
\begin{equation}
Z_{\rm whole} (\beta) = Tr \exp (-\beta \hat H_{\rm whole}) \ \ .
\label{Zext}
\end{equation}

During the recent years there has been increasing evidence that the mass
spectrum of the black hole spacetime might be discrete\cite{24}. If we denote these 
discrete mass eigenvalues of the mass operator ${\hat m} ={\hat H_{\rm whole}}$
by $m_n$ $(n=0,1,2,\dots)$ and the corresponding eigenvectors 
$\vert m_n \rangle$, we obtain an eigenvalue equation
\begin{equation}
{\hat H_{\rm whole}}\vert m_n \rangle={\hat m}\vert m_n \rangle=m_n\vert m_n \rangle \ \ .
\label{eigeneq}
\end{equation}
When the discrete energy spectrum is employed, the partition function (\ref{Zext}) becomes
\begin{equation}
Z_{\rm whole} (\beta) = \sum_{n=0}^{\infty} \langle m_n\vert \exp (-\beta {\hat m}) \vert m_n \rangle
                      = \sum_{n=0}^{\infty} \exp (-\beta m_n) \ \ .
\label{Zwhole}
\end{equation}
Since the Bekenstein-Hawking entropy of black holes is
\begin{equation}
S_{\rm BH} = \frac{1}{4}A \ \ ,
\label{BHentropy}
\end{equation}
where $A$ is the area of the event horizon, it is natural to assume an $\exp(\frac{1}{4}A)$-fold 
degeneracy in the possible mass eigenvalues $m_n$ of the hole. This assumption of degeneracy 
is justified because entropy, in general, can be understood as a logarithm of the
number of microstates corresponding to the same macrostate. Since for a \S black hole with mass $m$,
$A = 16 \pi m^2$, we are prompted to define
$g(m_n)$ as the number of degenerate states corresponding to the same mass eigenvalue 
$m_n$ such that
\begin{equation}
{\rm g}(m_n) = \exp (4\pi m_{n}^{2}) \ \ .
\label{degen}
\end{equation}
Hence, when the summation is performed over different mass eigenvalues only, we get for the whole maximally extended 
\S spacetime the partition function which takes the following form:
\begin{eqnarray}
Z_{\rm whole} (\beta) &=& \sum_{n=0}^{\infty} {\rm g} (m_n) \exp (-\beta m_n) 
\nonumber \\ 
																					 &=& \sum_{n=0}^{\infty} \exp (-\beta m_n + 4\pi m_{n}^{2} ) \ \ .
\label{Zwhole2}
\end{eqnarray}
Before investigating the partition function (\ref{Zwhole2}) any further, 
let us, at this point, turn our attention to the partition function corresponding to the exterior 
region of the \S black hole spacetime. 

Classically, the Hamiltonian $H_{\rm ext}$ of the exterior region of the \S black hole 
spacetime may be understood, in a certain foliation, as the total energy of the exterior region of the 
hole, although according to Bose et al. $H_{\rm ext}$ is the free energy of the whole black hole 
spacetime\cite{25}. To obtain the corresponding partition function for the exterior region, we replace, 
as before, the operator $\hat H$ of Eq. (\ref{defpart}) by an operator counterpart ${\hat H}_{\rm ext}$
of $H_{\rm ext}$ and we require, as before, that the mass spectrum is 
discrete. In contrast to our discussion concerning the partition function of the whole spacetime,
however, we assume the mass eigenstates to be non-degenerate. As a consequence, we get for the 
exterior region the partition function
\begin{eqnarray}
Z_{\rm ext}(\beta) &=&
\sum_{n=0}^{\infty}
\left\langle m_n \vert
\exp [-\beta (\hat m N_+ -2\hat m^2 N_0 )]
\vert m_n \right\rangle 
\nonumber \\
&=& \sum_{n=0}^{\infty}
\exp [-\beta ( m_n N_+ -2 m^2_n N_0 )] \ \ .
\label{Zext}
\end{eqnarray}
Note that the partition function (\ref{Zext}) is observer-dependent. To choose the same observer at
the asymptotic infinity as in Eq. (\ref{Zwhole2}) we must, again, fix the value of the lapse 
function at the right-hand-side spatial infinity such that $N_+ \equiv 1$.
From the point of view of such an observer, the partition function of the exterior region of the
\S black hole therefore takes the form:
\begin{eqnarray}
Z_{\rm ext}(\beta) = \sum_{n=0}^{\infty}
\exp [-\beta ( m_n  -2 m^2_n N_0 )] \ \ .
\label{Zext1}
\end{eqnarray}

To calculate the partition functions (\ref{Zwhole2}) and (\ref{Zext1}), 
we must assume, in addition, a specific spectrum for the  mass eigenvalues $m_n$ 
of the hole. In 1974 J. Bekenstein made a proposal, since then revived by several 
authors, that the possible eigenvalues of the area of the event horizon of the 
black hole are of the form\cite{26}:
\begin{equation}
A_n = \gamma n l^2_{\rm Pl} \ \ ,
\label{Bprop}
\end{equation}
where $\gamma$ is a pure number of order one, $n$ ranges over all non-negative 
integers, and $l_{\rm Pl} := (\hbar G/c^3)^{1/2}$ is the Planck length. When imposing
this proposal, we find that the partition function of the whole \S black hole spacetime 
is
\begin{equation}
Z_{\rm whole} (\beta) = \sum_{n=0}^{\infty} \exp 
\left(  -\frac{\beta}{4}\sqrt{\frac{\gamma n}{\pi}} + \frac{\gamma n}{4}\right)\ \ ,
\label{Zwhole3}
\end{equation}
and the partition function of the exterior region of the \S black hole is
\begin{equation}
Z_{\rm ext} (\beta ) = \sum_{n=0}^{\infty}\exp \left[ -\beta \left(
\frac{1}{4}\sqrt{\frac{\gamma n}{\pi}}-\frac{N_0}{8}\frac{\gamma n}{\pi}
\right) \right] \ \ ,
\label{Zext2}
\end{equation}
which both diverge very badly, indeed.

To actually calculate the partition functions (\ref{Zwhole3}) and (\ref{Zext2}), we have to deal
with the problem of diverging partition functions. Kastrup has suggested some very original and interesting solutions to the divergency
problem\cite{11}. Our solution to the problem of a diverging partition function in the case of the whole 
maximally extended \S spacetime is to study not the partition function of the whole spacetime itself 
but, instead, the partition function of the {\it radiation} emitted by the hole. When obtaining the 
partition function for the radiation, we assume that the evaporation of the hole is a reversible
process. In other words, we assume that the entropy of the hole is converted exactly into the entropy
of the radiation. A validity of this assumption has been investigated by Zurek\cite{27}. His conclusion was that if 
the temperature of the heat bath is the same as that of the hole, then the black hole evaporation is a reversible 
process.

First, we choose the zero point of the energy emitted by the hole. This could be
done in many ways, but we choose the total energy of the radiation emitted to be 
zero when the hole has evaporated completely leaving nothing but radiation. With 
this choice of the zero point of the total energy of the radiation, we find that
the relationship between the energy $E^{\rm rad}$ emitted by
the hole and the mass $m$ of the \S black hole measured at the asymptotic 
right-hand-side infinity is
\begin{equation}
E^{\rm rad} = - m  \ \ .
\label{Erad}
\end{equation}

If all the entropy of the hole is converted into the entropy of the radiation by means of
transitions between degenerate black hole energy eigenstates, then the radiated energy spectrum
is degenerate, too, and the number of the degenerate states corresponding to the same total energy 
emitted by the hole since its formation up to the point where the \S mass has achieved the value 
$m_n$, is given by a function $g^{\rm rad}(m_n)$. It is fairly obvious that $g^{\rm rad}(m_n)$
increases when $m_n$ decreases. In an ideal case, all the entropy of the hole is exactly converted
to the entropy of the radiation. In that case we may choose
\begin{equation}
g^{\rm rad}(m_n) = \exp(\frac{1}{4}A_0-4\pi m_n^2) \ \ ,
\label{grad}
\end{equation}
where $A_0$ is the initial surface area of the black hole horizon, measured just before the hole 
has begun its evaporation.  
In other words, the decrease of the black hole entropy from $\frac{1}{4}A$ to
$\frac{1}{4}(A-dA)$ increases the number of degenerate states of the radiation
emitted by the hole by a factor $\exp (\frac{1}{4}dA)$. This choice reflects 
the fact that just after the hole has been formed, and not radiated yet, the
entropy of the radiation is zero, whereas the entropy is $\frac{1}{4}A$ after the 
hole has evaporated completely.

Now, since $E^{\rm rad}=-m$ and $H_{\rm whole}=m$, we argue that
\begin{equation}
H_{\rm whole}^{\rm rad}= -m \ \ .
\label{Hradwhole}
\end{equation}
To obtain the partition function for the radiation of the whole \S spacetime, Eqs. (\ref{defpart}),
(\ref{eigeneq}), (\ref{grad}) and (\ref{Hradwhole}) yield:
\begin{eqnarray}
Z_{\rm whole}^{\rm rad} (\beta) &=&\sum_{n=0}^{\infty} {\rm g^{rad}} (m_n) \exp (\beta m_n) 
\nonumber \\ 
&=&\exp \left(\frac{1}{4}A_0\right)\sum_{n=0}^{\infty} \exp (\beta m_n-4\pi m_n^2) \ \ .
\label{Zwholerad}
\end{eqnarray}
When Bekenstein's proposal (\ref{Bprop}) is used, we get a partition function
\begin{equation}
Z_{\rm whole}^{\rm rad} (\beta) = \exp \left(\frac{1}{4}A_0\right)\sum_{n=0}^{\infty} \exp 
\left(  \frac{\beta}{4}\sqrt{\frac{\gamma n}{\pi}} - \frac{\gamma n}{4}\right)
\label{Zwholerad2}
\end{equation}
describing the radiation emitted by the \S black hole. It is easy to see that 
$Z_{\rm whole}^{\rm rad}$ converges very nicely.

In comparison, let us obtain, by means of the same procedure as above, the partition function of the 
radiation emitted by the spacetime exterior to the \S black hole. We choose the zero 
point of the energy of the radiation emitted by the external spacetime in the same way as before. 
This radiational energy should be understood as arising from
transitions between the energy states of the gravitational field corresponding to the exterior region 
of the \S
black hole spacetime. The zero points of energy emitted by both the Kruskal and the exterior spacetime
can be chosen to coincide because the distant observer outside the hole observes the same 
energy $E^{\rm rad}$. 

Now, since all the energy of the exterior region is assumed to be converted into the energy of the 
radiation, the Hamiltonian of the radiation of the exterior region may then be taken to be
\begin{equation}
H_{\rm ext}^{\rm rad} = - H_{\rm ext} \ \ .
\label{Hextrad}
\end{equation}
To obtain the partition function $Z_{\rm ext}^{\rm rad}$ one uses Eqs. (\ref{defpart}), (\ref{eigeneq}) and 
(\ref{Hextrad}). These equations give a partition function
\begin{equation}
Z_{\rm ext}^{\rm rad} (\beta) = \exp \left(\frac{1}{4}A_0\right)\sum_{n=0}^{\infty} \exp (\beta m_n-2N_0m_n^2) \ \ ,
\label{Zextrad}
\end{equation}
where we chose an appropriate normalization constant to the partition function. This is 
allowed, since the normalization does not have any effect on the measurable thermodynamical quantities, like 
temperature, of the system.

Applying, again, Bekenstein's proposal (\ref{Bprop}) to Eq. (\ref{Zextrad}), we get
\begin{equation}
Z_{\rm ext}^{\rm rad} (\beta) = \exp \left(\frac{1}{4}A_0\right) \sum_{n=0}^{\infty} \exp 
\left[ 
\frac{\beta}{4}
\left( 
\sqrt{\frac{\gamma n}{\pi}} -\frac{N_0}{2}\frac{\gamma n}{\pi} 
\right) 
\right] \ \ ,
\label{Zextrad2}
\end{equation}
which, when keeping $N_0$ fixed, converges, too.

Let us next calculate the converging partition functions (\ref{Zwholerad2}) and (\ref{Zextrad2}).  
Assuming that $\beta$ is very big, we may approximate the sums (\ref{Zwholerad2}) and (\ref{Zextrad2})
by integrals\cite{28}:
\begin{mathletters}
\label{intZ}
\begin{eqnarray}
Z_{\rm whole}^{\rm rad} (\beta)
& \approx & 
\exp \left(\frac{1}{4}A_0\right)\int_0^{\infty} dn \exp \left(  \frac{\beta}{4}\sqrt{\frac{\gamma n}{\pi}} - \frac{\gamma n}{4}\right)
\nonumber \\
& = & 
\exp \left(\frac{1}{4}A_0\right)\left[\frac{4}{\gamma} +\frac{\beta}{\gamma}
\left[ 
1+{\rm erf}\left( \frac{\beta}{4\sqrt{\pi}}\right)
\right]
\exp \left( \frac{\beta ^2}{16\pi}\right)\right] 
\ \ , 
\\ 
\label{intZwholerad} 
Z_{\rm ext \ \ }^{\rm rad} (\beta) 
& \approx & 
\exp \left(\frac{1}{4}A_0\right)\int_{0}^{\infty} dn \exp 
\left[ 
\frac{\beta}{4}
\left( 
\sqrt{\frac{\gamma n}{\pi}} -\frac{N_0}{2}\frac{\gamma n}{\pi} 
\right) 
\right] \nonumber \\
& = &
\exp \left(\frac{1}{4}A_0\right)\left[\frac{8\pi}{\gamma \beta N_0} + \frac{4}{\gamma}\sqrt{\frac{2}{\beta}}
\left( \frac{\pi}{N_0} \right) ^{3/2} 
\left[ 
\frac{1}{2} +\frac{1}{2}{\rm erf}\left( \frac{\beta}{8N_0}\right) ^{1/2}
\right]
\exp \left( \frac{\beta}{8N_0}\right)\right] \ \ ,
\label{intZextrad}
\end{eqnarray}
\end{mathletters}
where ${\rm erf}(x)$ is the error function. 

If we, now, choose
\begin{equation}
N_0=\frac{2\pi}{\beta} \ \ ,
\label{N0}
\end{equation}
then 
\begin{equation}
Z_{\rm whole}^{\rm rad} = Z_{\rm ext}^{\rm rad} := Z^{\rm rad} \ \ .
\label{Zs}
\end{equation}
This is the main result of this paper.  It should be noted that this result in not just an artefact of an
approximation of a sum by an integral, but it holds even for exact expressions (\ref{Zwholerad2}) and 
(\ref{Zextrad2}). We shall discuss the consequences of our result at the end of this section. Let us, 
in the meantime, try to justify Eq. (\ref{N0}).

If Eq. (\ref{N0}) holds, then Eqs. (\ref{intZ}) give the semiclassical
partition function of the radiation observed by an external observer at asymptotic 
infinity:
\begin{equation}
Z^{\rm rad}(\beta ) \approx \exp \left(\frac{1}{4}A_0\right)
\frac{2\beta }{\gamma }\exp \left( \frac{\beta ^2}{16\pi} \right) \ \ .
\label{Zrad}
\end{equation}
It is easy to show that the upper bound for the absolute error made, when replacing the sums (\ref{Zwholerad2}) and 
(\ref{Zextrad2}) by integrals (\ref{intZ}) is, in the leading order approximation, 
$\exp (1/4 A_0+\beta^2 /16\pi)$. If one compares the result (\ref{Zrad}) to the absolute error made when
replacing the sums by integrals, one notices that, for very big $\beta$, the fractional error is much 
smaller than unity. 
Hence, in the highest order approximation, the resulting partition function (\ref{Zrad}) 
approximates the sums (\ref{Zwholerad2}) and (\ref{Zextrad2}) very well and, most importantly, the 
effect of the error bars on the thermodynamical quantities is negligibly small. 

We now require that the energy expectation value of the radiation is: 
\begin{equation}
\langle E^{\rm rad} \rangle := -\frac{\partial}{\partial \beta} 
\ln Z^{\rm rad}_{\rm ext} (\beta )  = -\langle m \rangle \ \ .
\label{expErad}
\end{equation}
When $\beta$ and $\langle m \rangle$ are taken to be  very big, we get from (\ref{expErad}):
\begin{equation}
-\frac{\beta}{8\pi} +{\cal O}(\beta^{-1}) = - \langle m \rangle \ \ ,
\label{beta1}
\end{equation}
which, in turn, is the same as
\begin{equation}
\beta \approx 8\pi\langle m \rangle  \ \ .
\label{beta2}
\end{equation}
This, on the other hand, corresponds to the choice
\begin{equation}
N_0 \approx \frac{1}{4\langle m \rangle } \ \ .
\label{N02}
\end{equation}
It was noted by Bose et al. that when Einstein's field equations are satisfied, the quantity
$N_0$ can be expressed as $N_0 = \kappa \frac{dT}{dt}$\cite{25}, where $T$ is the \S time coordinate,
i.e. the Killing time, $t$ is the global time coordinate, and $\kappa = \frac{1}{4m}$ is the 
surface gravity of the black hole. Now, Eq. (\ref{N02}) implies that, in the semiclassical limit,
$\frac{dT}{dt}=1$, which states that the time coordinate $t$ equals with the \S time $T$. In other words, the
meaning of the choice (\ref{N0}) is that the spacetime foliation near the horizon of the \S
black hole is determined by the \S time coordinate $T$. Since the \S time coordinate is just the time coordinate 
used by our external observer at rest when he makes observations on the spacetime properties, one may regard
the choice (\ref{N0}) justified on grounds of our aim to describe the black hole thermodynamics from the point of
view of a faraway observer at rest. On the other hand, if one requires that, in the leading approximation,
$N_0 \approx \frac{1}{4\langle m \rangle}$, and that $-\frac{\partial}{\partial \beta} 
\ln Z^{\rm rad}_{\rm ext} (\beta ) = -\langle m \rangle$, then -- as noted in Ref.\cite{25} -- one gets $\beta \approx 
4C\langle m \rangle$, which gives $N_0 \approx \frac{C}{\beta}$, where the constant $C$ can be chosen to be 
$2\pi$. Hence, if we use a \S-type foliation right from the beginning, we can obtain, up to a constant, the
choice  (\ref{N0}).

It is well known that the entropy $S$ of any thermodynamical system, described by a partition
function $Z$, can be calculated from an expression
\begin{equation}
S=\ln Z -\beta \frac{\partial}{\partial \beta}\ln Z \ \ .
\label{entropy}
\end{equation}
When substituting $Z^{\rm rad}$ into Eq. (\ref{entropy}), one gets an approximation to
the entropy of the black hole radiation:
\begin{equation}
S^{\rm rad} = \frac{1}{4}(A_0 - A)  + \frac{1}{2}\ln A + \ln \left(\frac{4\sqrt{\pi}}{\gamma}\right) -1 +
{\cal O}\left( A^{-1/2}\right)\exp \left( -\frac{1}{4}A\right) \ \ .
\label{Sextrad}
\end{equation}
Hence, when the area of the black hole has shrinked from $A_0$ to $A$, 
the entropy carried away by the radiation is, in the leading order approximation, $\frac{1}{4}(A_0-A)$. 
Under assumption that the black hole radiation is a reversible process, this result is compatible with the 
Bekenstein-Hawking expression for black hole entropy: A decrease of the area by an amount $A_0-A$ decreases
the entropy of the hole by an amount $\frac{1}{4}(A_0-A)$. The error made when approximating the sum by an 
integral causes an error in the entropy which is of order ${\cal O}\left( A^{-1/2}\right)$.

We have obtained two partition functions $Z_{\rm whole}^{\rm rad}$ and 
$Z_{\rm ext}^{\rm rad}$. When obtaining the partition function $Z_{\rm ext}^{\rm rad}$
for the radiation emitted by the exterior region of the hole, the mass eigenstates were assumed to be discrete --
as proposed by Bekenstein -- and non-degenerate. When obtaining the partition function 
$Z_{\rm whole}^{\rm rad}$ of the radiation emitted by the whole Kruskal spacetime, however, we had to make an {\it ad hoc} assumption of an 
$\exp (\frac{1}{4}A)$-fold degeneracy in the discrete mass eigenstates of the hole to get 
the correct black hole entropy. Still, the two partition functions turned out to be exactly the same 
from the point of view of a distant observer. This is a very interesting result. Does it bear any 
implications relevant to the question of the nature of the black hole entropy?

Our investigation suggests two possible interpretations to the black hole entropy. The first 
interpretation is that the entropy of the hole is simply caused by the fact that an
external observer cannot make any observations on the interior region of the black hole. As a
consequence, the physics of a black hole is physics of its external region for such an observer, and it is sufficient to 
consider the statistical mechanics of that external region only. This interpretation is supported
by our straightforward calculation which gives correctly the Bekenstein-Hawking entropy, without 
assuming any degeneracy in the mass eigenstates.

Another interpretation is more conservative: The entropy of the hole 
is interpreted as a huge degeneracy in the mass eigenstates of the whole black hole spacetime -- including
the interior region of the hole. When using this interpretation to obtain the Bekenstein-Hawking 
entropy one must make an {\it ad hoc} assumption about a vast $\exp (\frac{1}{4}A)$-fold degeneracy 
in the mass eigenstates.

What, then, are pro's and con's of the two viewpoints? From the first point of
view, let us call it as an external point of view, the degrees of freedom of the collapsing matter, except
the mass, are completely lost. Thus, the external point of view indicates that the information
contained in the collapsing matter is not just hovering at some place, but completely and totally lost, 
whereas the conventional viewpoint somehow allows one to include the information about the 
degrees of freedom of the collapsing matter into the microstates of the hole itself. The loss of 
information, as known, leads to severe fundamental problems. These problems are
discussed, for example, in\cite{29,30,31,32,33,34,35,36}. On the other hand, the external view makes it possible to consider \S
black holes as objects having one physical degree of freedom only. This feature of the external point of 
view makes it appealing to us, as it -- unlike the conventional point of view -- is in perfect harmony
with the no-hair theorem. Hence, one does not necessarily need to be concerned with how the quantization
itself might bring along a vast number of additional degrees of freedom.

\section{Conclusion}
\label{sec:conc}

In this paper we have obtained the partition function of the \S black hole by means of 
two different Hamiltonians $H_{\rm whole}$ and $H_{\rm ext}$. These Hamiltonians 
describe, respectively, the whole maximally extended \S spacetime, and the exterior region of the 
\S black hole. The whole Hamiltonian thermodynamics was considered in Lorentzian 
spacetime. The main reason for not producing a euclideanized partition function
of the \S black hole was that we wanted to include the interior of the black
hole in the analysis. After writing the Hamiltonians, we obtained the corresponding
partition functions which can be viewed, respectively, as the partition functions of the whole 
maximally extended \S spacetime, and the spacetime region exterior to the black hole, from the point
of view of a faraway observer at rest. 

We found that these two partition functions coincide. To obtain this result, however, we were compelled to
assume an $\exp (\frac{1}{4}A)$-fold degeneracy in the mass eigenstates when calculating the partition function
of the whole spacetime, whereas no degeneracy was needed to be assumed when calculating the partition 
function for the exterior region. In addition, we chose the spacetime foliation near the horizon of the \S
black hole to be determined by the \S time coordinate $T$ which fixed, up to a constant, the quantity $N_0$.

To check the correctness of our partition functions, we used Bekenstein's proposal for a discrete 
area spectrum of black holes to calculate the Bekenstein-Hawking entropy. Unfortunately, the partition 
functions of the whole black hole spacetime, and the spacetime region exterior to the hole were found to 
diverge, but we managed to solve the divergency problem, however, by turning our attention to the 
{\it radiation} emitted by the hole. More precisely, we obtained the partition functions of radiation
emitted when either the whole black hole spacetime or its exterior region are assumed to perform
transitions from a one state to another. When obtaining the partition functions of radiation we 
assumed that the evaporation of the hole is a reversible process and that all the energy and the entropy 
of the hole are exactly converted to the energy and the entropy of radiation. The resulting partition
functions for radiation were found to converge very nicely producing, in the leading order 
approximation, the Bekenstein-Hawking entropy of black holes.  

Our investigation suggested that the black hole entropy can be interpreted in two possible ways. First, 
there is the conservative view that the entropy of black holes may be understood as a result of a huge
degeneracy in the mass eigenstates of the whole black hole spacetime. The degeneracy of the eigenstates
might somehow, in a still unexplained manner, allow one to include the degrees of freedom of the collapsed 
matter, but the view is in contradiction with the no-hair theorem. The second view -- called the external 
point of view -- is that the entropy of black holes is, quite simply, caused by the fact that the interior
region of black hole spacetime is separated from its exterior region by a horizon. Because of that, one
might be justified to take a view that black hole statistical mechanics is, for an external observer, 
statistical mechanics of its exterior region. This point of view allows one to obtain the Bekenstein-Hawking
entropy without assuming any degeneracy in the mass eigenstates of the hole. The result is in harmony
with the no-hair theorem, but allows a complete loss of information, since the degrees of freedom of
the matter, except the total mass $M$, have vanished. We have thus two complementary points of view to the 
interpretation of black hole entropy, of which neither is quite completely satisfactory: The conservative
view is in conflict with the no-hair theorem, whereas the external point of view, although it is physically
appealing and in harmony with the no-hair theorem, implies a tremendous loss of information. It remains
to be seen whether these two possible interpretations could somehow be unified into a single, consistent
description of black holes. 

\acknowledgments
We are grateful to Jorma Louko and Markku Lehto for their constructive criticism during the preparation of this
paper. P. R. was supported by the Finnish Cultural Foundation, Wihuri Foundation and Nyyss\"onen Foundation.

\end{document}